\documentclass[10pt]{article}
\usepackage[explicit]{titlesec}
\usepackage{hyphenat}
\usepackage{ragged2e}
\RaggedRight

\usepackage{soul}
\setul{.6pt}{.4pt}

\usepackage{geometry}
\usepackage{fancyhdr}
\usepackage[labelfont={footnotesize,bf} , textfont=footnotesize]{caption}
\titleformat{\section}
  {\normalfont}{\thesection}{1em}{\MakeUppercase{\textbf{#1}}}
\titlespacing\section{0pt}{0pt}{-10pt}
\titleformat{\subsection}
  {\normalfont}{\thesubsection}{1em}{\textit{#1}}
\titlespacing\subsection{0pt}{0pt}{-8pt}

\makeatletter
\newcommand\sixteen{\@setfontsize\sixteen{17pt}{6}}
\renewcommand{\maketitle}{\bgroup\setlength{\parindent}{0pt}
\begin{flushleft}
\sixteen\bfseries \@title
\medskip
\end{flushleft}
\textit{\@author}
\egroup}
\makeatother

\usepackage[sort&compress]{natbib}
\makeatletter
\renewcommand\@biblabel[1]{\textbf{#1.}\hfill}
\makeatother

\bibpunct{}{}{,~}{s}{,}{,}
\usepackage{ragged2e}
\usepackage{wrapfig}
\usepackage{algorithm}
\usepackage{algorithmic}
\usepackage{mathtools}
\usepackage{amsmath}

\DeclareMathOperator*{\minimize}{minimize}

\title{Neighbourhood Evaluation Criteria for Vertex Cover Problem}

\author{
Kaustubh K Joshi\\ \medskip 
E-mail: kaustubhkj@gmail.com
}
\begin{document}

\vspace*{.01 in}
\maketitle
\vspace{.12 in}

\begin{abstract}
\justifying
The Vertex Cover Problem is a well-known graph-oriented NP-complete problem. This paper attempts to construct an approximate polynomial time algorithm using the Neighbourhood Evaluation Criteria (NEC) for finding near-optimal solutions.
A degree count is kept in check for each vertex and the highest count based vertex is included in our cover set. In the case of
multiple equivalent vertices, the one with the lowest neighbourhood influence is selected. In the case of still existing multiple
equivalent vertices, the one with the lowest remaining active vertex count (the highest Independent Set enabling count) is
selected as a tie-breaker. Algorithm specifications and results have been stated ahead.
\end{abstract}

\section*{keywords} 
Vertex Cover Problem, Approximation Algorithm, NP-Complete Problem, Heuristic Technique 

\vspace{.12 in}


\section{Introduction}
The Vertex Cover Problem is a topic of graph theory and combinatorics and is classified under the NP-complete category. This problem often attracts a lot of research and practitioners due to its vast application in the real world (telecommunication [1] , electrical engineer,etc.) and for it's ability of being a complement to the Independent Set Problem which can help in solving SAT problems. This in term helps to solve various other NP problems (Circuit SAT, Max Clique selection [10],etc.) which can be reduced to a SAT problem.
\\
As stated by Richard Karp [2] the selection of optimal nodes for a Vertex Cover in a graph is an NP-complete problem. Hence a pure Polynomial run time algorithm with always exact solution is currently improbable.
The simplest algorithm takes into account insertion of vertices via random edge selection, however this leads to a vertex cover size (worst case) of two times the optimal value ($1+\epsilon = 2$). Most heuristic/approximate algorithms have an optimal selection ratio ( approximate vertex cover : optimal vertex cover) of $1 + \epsilon$ where $\epsilon \in [0,1]$ .
\\ 
Modern approaches[3][4] to solving the Vertex Cover Problem utilizes Machine Learning techniques or meta heuristic algorithms[11] to obtain approximate results, these techniques attempt to search for patterns and try to generalize them to other sets.This approach often fails as some graphs are designed in a counter-intuitive to break such generalizations. Approximate algorithm based approaches utilize a greedy method (using a support count/degree count/weighted distribution) to drive the selection criteria. While this works for basic cases it also leads to higher selection ratio in nuanced cases. As stated by Dinur and Safra[5],it is an NP-Hard task to get $1+\epsilon < 1.3606$ bound for all instances.

\pagebreak

\section{Problem Description}

Given an undirected graph G(V,E) let C = \{ u : $\forall e$ ( $e\in E$), $\exists u$ ( $u \in V$) incident to e\} . This set C is known as a vertex cover. Our problem is to find the smallest vertex cover size |C| satisfying the above statement.

As with most methods of approach to NP problems[6], there exist two types of solving methods. The verification method checks whether our candidate solution is in fact a valid solution and then continue to tighten the constraints iteratively. The optimization method finds the exact solution by optimizing a given function. For verification cases the problem can be expressed as follows : 
\begin{center}
$\bigwedge_{(v_{i},v_{j})\in E}$  ($v_{i} \vee v_{j}$)
\end{center}

This translates into a 2-SAT type of problems that can be solved using implications graphs and verified quickly due to the property of it being a subclass of SAT problems. For optimization cases the problem can be expressed as follows : 
\begin{center}
$\displaystyle{\minimize_{v\in C*} (\lambda)e^{(E\cdot C-E\cdot C^{\star})} + (\beta)|C-C^{\star}|}$ (where $E = Z_{m \times n}^{\{0,1\}}$ ; $C,C^{\star} = Z_{n \times 1}^{\{0,1\}}$)
\end{center}

Here E acts as an Adjacency Matrix and C,C* acts as selection vectors. C is our optimal solution and C* is our candidate solution. We can alter the value of $\lambda$ parameter to ensure that C* doesn't try to undercut C. Similarly we can alter the $\beta$ parameter to ensure that C* does not overshoot the desired selection.

\section{Examples}

\subsection{Bipartite Graphs}

Consider a Bipartite Graph of type K(m,n) . In such a case the NEC algorithm will pick up nodes from the smaller set since these nodes will have higher degree count compared to the larger set. 
\\
In this example we can see the graph as being the a join operation between set \{A,B,C,D,E\} and \{F,G,H\}. Since a smaller set will have a larger degree value; this leads to our minimum vertex cover to be composed of \{F,G,H\}.

\begin{figure}[ht!]
\centering
\includegraphics[width=5cm,height=5cm]{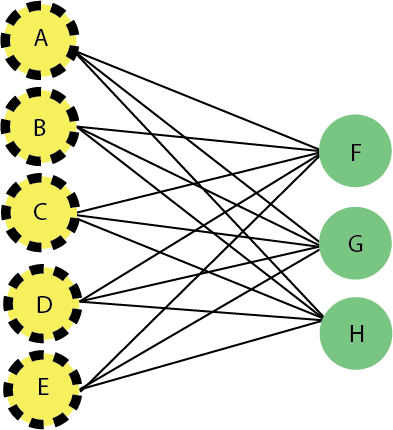}
\caption{Bipartite Graph of type K(5,3)}\label{bp graph}
\end{figure}

\subsection{Complete Graphs}

Consider a Complete Graph of size-k. In such a case the NEC algorithm will pick up any node since all of them are equivalent and do not have any tiebreaker for providing a benefit for selection. This leads our graph to now become of type k-1. Inductively our graph will finally become a complete graph of size 2 (which leads to a single node selection and causes only one node to be left unselected). Hence our optimal cover set will be of size k-1. 
\\
In this example we are provided with a K-5 graph. As stated above we will eventually end up with a single non-selected node and our minimum vertex cover will be comprised of \{B,C,D,E\}

\begin{figure}[ht!]
\centering
\includegraphics[width=5cm,height=5cm]{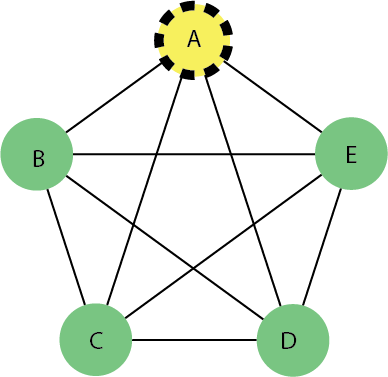}
\caption{A Complete Graph of type K5}\label{k5 graph}
\end{figure}

\subsection{Random Graphs}

Consider a Random Graph of type G(V,E). In such a case there exist multiple selection choices. Since the initial selection of NEC is greedy, we will choose a candidate node with highest degree. In case of multiple nodes with highest degree, we will use our tie breaker rule of minimum neighbourhood degree and lowest facet selection to choose our optimal candidate node.
\\
In the give example we are provided with a G(14,16) random graph. Utilizing the algorithm, we will first end up selecting H since it has the highest degree. After that we will select C, then B and so an and so forth until we end up with our minimum vertex cover of \{B,C,G,H,K,M\}

\begin{figure}[ht!]
\centering
\includegraphics[width=12cm,height=6cm]{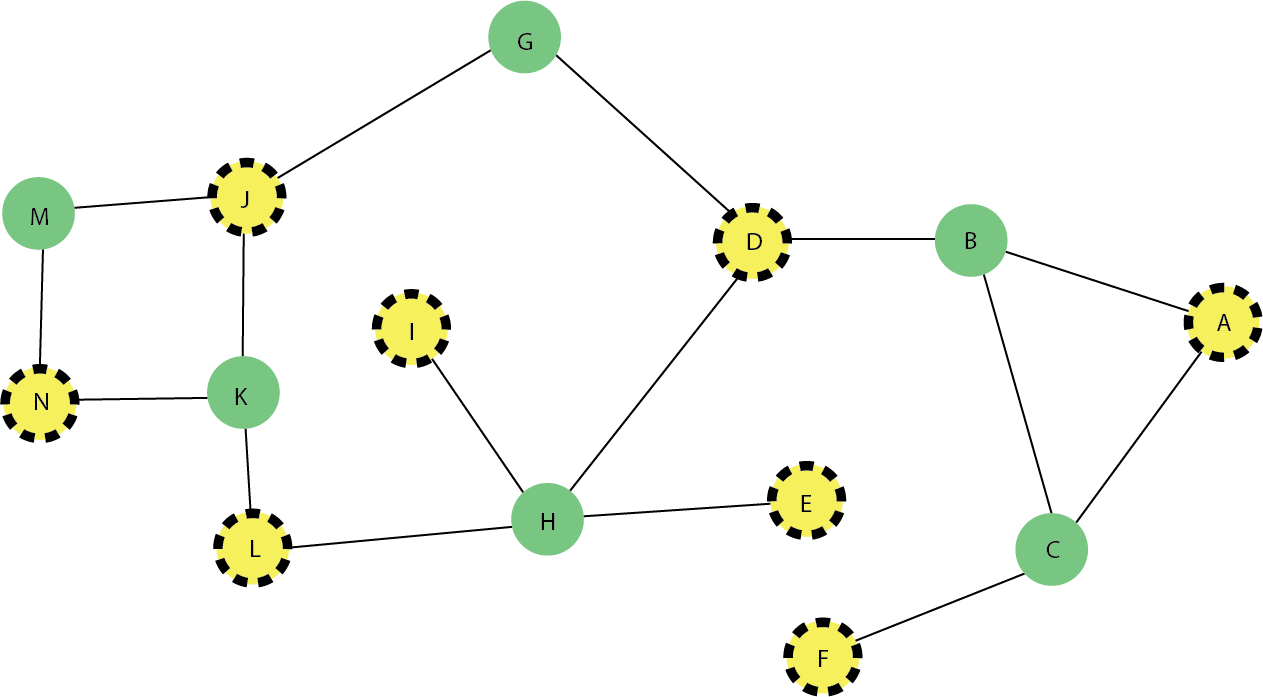}
\caption{A Random Graph}\label{random graph}
\end{figure}

\pagebreak

\section{Neighbourhood Evaluation Criteria (NEC) }

The idea behind NEC is that the nodes with highest degree counts are of interest to use while doing selection since they are connected to a variety of other nodes.However a straightforward approach won't work for clashes. In the case of multiple clashing nodes, we would like to select a node which has the lowest neighbourhood degree count since this would indirectly imply that it's connected to nodes in a lower facet hence removing it's need for inclusion in our optimal candidate selection set. In the case of further existing ties, we utilize a tie breaker that checks the strength of the selected node vs the competitor node by selecting the node which will lead to the exclusion of the most single degree nodes.

\begin{algorithm}
\caption{Neighbourhood Evaluation Criteria}
\begin{algorithmic}
\STATE $V \leftarrow$ Set of all vertices in the graph
\STATE $E \leftarrow$ Set of all edges in the graph
\STATE $A[V] \coloneqq$ Adjacency List Array
\STATE $N[V] \coloneqq$ Degree count Array
\STATE $C^{*} \coloneqq$ Empty set for selected vertices
\STATE $E^{*} \coloneqq$ Empty set for covered edges
\STATE $isActive[V] \coloneqq$ Boolean array assigned to true
\WHILE{$|E^{*}| < |E|$}
\STATE {$candidateNode \leftarrow \phi$}
\FOR{$u \in V$}
\IF{isActive[u] is true and N[u] > N[candidateNode] }
\STATE {$candidateNode \leftarrow$ u}
\ENDIF
\ENDFOR
\FOR{$u \in V$ and $u \neq candidateNode$}
\IF{ $\sum_{j \in A[u]} (isActive[j]*N[j])$ < $\sum_{k \in A[candidateNode]} (isActive[k]*N[k])$ }
\STATE {$candidateNode \leftarrow$ u}
\ELSIF{$\sum_{j \in A[u]} (isActive[j]*N[j])$ == $\sum_{k \in A[candidateNode]} (isActive[k]*N[k])$}
\IF {$\sum_{j \in A[u]} (isActive[j]*\{N[j]==1\})$ > $\sum_{k \in A[candidateNode]} (isActive[k]*\{N[k]==1\})$ }
\STATE {$candidateNode \leftarrow$ u}
\ENDIF
\ENDIF
\ENDFOR
\FOR{$u \in A[candidateNode]$}
\IF{isActive[u] is true  }
\STATE { $E^{*}\leftarrow E^{*}\cup$ (candidateNode,u)}
\STATE {$N[candidateNode] \leftarrow N[candidateNode]-1$ }
\IF{$N[u] == 0 $ }
\STATE {$isActive[u] \leftarrow$ false}
\ENDIF
\ENDIF
\ENDFOR
\STATE { $C^{*}\leftarrow C^{*}\cup$ candidateNode}
\STATE {$isActive[candidateNode] \leftarrow$ false}
\ENDWHILE
\end{algorithmic}
\end{algorithm}

\subsection{Computational Complexity}

Big O time complexity has been calculated as follows: The maximum number of times edge loop will take place will be k where k is the number of candidate vertices which cover all the edges. For each loop the following tasks are done: Selecting an optimal node and updating the values for said optimal node. Optimal node selection is an O($V^2$) time complexity task and value update is an O(V) time complexity task. Our overall complexity comes out to be O($kV^2$). A complete algorithm design will have a time complexity of O($kn + 1.29175^{k}k^{2}$) as stated by Rolf and Peter[12]. This leads us to have an approximate result algorithm which produces optimal or near optimal selection while ensuring fast asymptotic time complexity.

\subsection{Selection criteria}

Degree Count - Initial Selection : This is the greedy part of the algorithm which selects a candidate node based on degree count. Higher degree counts are selected over lower greedy count for the initial candidate node. This helps ensure that lower facet based nodes are not being given priority for selection (hence avoiding potential pitfalls)
\\
Lower Neighbourhood score - Tiebreaker : After the initial candidate node has been selected, there might still exist multiple nodes with the same degree count. To tackle this case we select the node which has the lower degree summation of its surrounding neighbours. This helps ensure that the potential node that we are taking into account doesn't end up disrupting future candidate nodes
\\
Highest deactivation score - Tiebreaker : In case of equal Neighbourhood scores a preference is taken for choosing a node which tends to cover the lowest faceted nodes. Hence a higher deactivation score is preferred as the final tiebreaker.

\section{Experimentation Results}

The proposed algorithm has been designed to run in C++11. Hardware specifications are as follows : intel Core i5-8300H 2.3Ghz CPU and 8 GBs of RAM. Experiments were carried out on the DIMACS dataset[8] for accuracy and verification. Stress testing of NEC was doing using complete graphs since these graphs have n-1 optimal nodes for satisfying vertex cover and requires complete search since tie breakers will not work in such graphs. 

\subsection{Complete Graph Performance}

The NEC algorithm is able to calculate the exact optimal count in O($kV^2$) time. This was verified by using polyfit(x,y,n) in OCTAVE. For n $\geq$ 3 the coefficients for x $\geq$ 2 were insignificant (near zero). The results have been plotted in Fig. 4. and tabulated in Table 1. This result has been performed purely for time complexity reasons since complete graphs by default will always produce k-1 vertices as a Minimum Vertex Cover size

\begin{figure}[htp!]
\centering
\includegraphics[width=12cm,height=6cm]{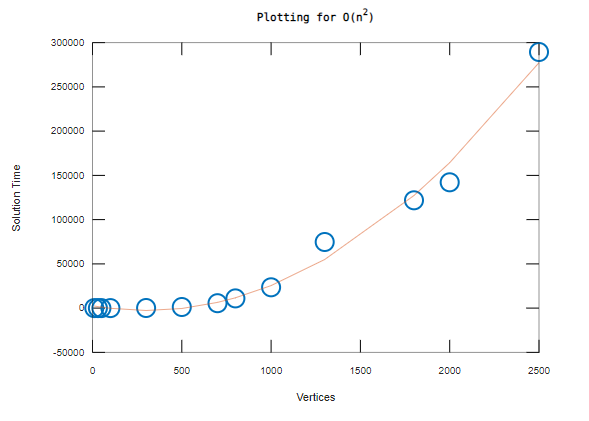}
\caption{Resultant plot for performance vs vertices}\label{evidence}
\end{figure}

\begin{table}[htp!]
\label{complete result table}
\begin{tabular}{ | c | c | c | c | c | c | c | c |}
\hline
 \textbf{Graph Type} & \textbf{Vertices} & \textbf{Optimal Vertices} & \textbf{Candidate Vertices}& \textbf{$1+\epsilon$} & \textbf{Time(milli-sec)}\\ 
\hline
\textbf{K-10} & 10 & 9 & 9 & 1 & 2\\
\hline
\textbf{K-30} & 30 & 29 & 29 & 1 & 3\\
\hline
\textbf{K-50} & 50 & 49 & 49 & 1 & 4\\
\hline
\textbf{K-100} & 100 & 99 & 99 & 1 & 7\\
\hline
\textbf{K-300} & 300 & 299 & 299 & 1 & 185\\
\hline
\textbf{K-500} & 500 & 499 & 499 & 1 & 1268\\
\hline
\textbf{K-700} & 700 & 699 & 699 & 1 & 5620\\
\hline
\textbf{K-800} & 800 & 799 & 799 & 1 & 10965\\
\hline
\textbf{K-1000} & 1000 & 999 & 999 & 1 & 23523\\
\hline
\textbf{K-1300} & 1300 & 1299 & 1299 & 1 & 74695\\
\hline
\textbf{K-1800} & 1300 & 1299 & 1299 & 1 & 121816\\
\hline
\textbf{K-2000} & 2000 & 1999 & 1999 & 1 & 142142\\
\hline
\textbf{K-2500} & 2500 & 2499 & 2499 & 1 & 289326\\
\hline
\end{tabular}
\caption{Results of NEC on Complete graph types }
\end{table}

\subsection{DIMACS performance}
While the algorithm displays an $\epsilon$ value of 0 for many graphs, in other cases the algorithm the algorithm is off by a few extra vertices (for e.g.: in case of Sanr400-0.5 NEC selects 3 extra vertices) and in worst case utilizes an $\epsilon$ value of 0.5 (meaning 50\% extra vertices can be picked up). The strength of the algorithm lies in the fact that it can produce near optimal minimum vertex cover sets while using an O($kV^{2}$) computation time and just O(E+V) memory space.
\\
When compared to COVER (a probabilistic Minimum Vertex Cover algorithm that selects an optimal solution via multiple iterations[9]) NEC achieves comparable results on a lot of graphs, however due to it's uniform random approach of selecting criteria satisfying nodes COVER can probabilistically achieve better results given multiple tries. NEC on the other hand is deterministic by nature and hence achieves consistent and faster results. For e.g. - For the graph Johnson32-2-4, COVER calculates the solution in 920 milliseconds whereas NEC calculates the solution in 56 milliseconds, hence having a 16 times faster performing speed. In another graph called Keller5, NEC includes 11 additional vertices while computing these results 50 times faster than COVER. Results have been tabulated in Table 2.

\begin{table}[htp!]
\centering
\begin{tabular}{ | c | c | c | c | c | c | c | c |}
\hline
 \textbf{Graph Type} & \textbf{|V|} & \textbf{|C|} & \textbf{|C*|}& \textbf{$1+\epsilon$} & \textbf{Time(milli-sec)} & \textbf{COVER |C*|} & \textbf{COVER Time(avg)}\\ 
\hline 
\textbf{Brock800\_1} & 800 & 777 & 782 & 1.006 & 43 & 782 & 40512\\
\hline 
\textbf{Brock800\_2} & 800 & 776 & 784 & 1.103 & 48 & 780 & NA \\
\hline 
\textbf{Brock800\_3} & 800 & 775 & 782 & 1.009 & 52 & 780 & NA\\
\hline 
\textbf{Brock800\_4} & 800 & 774 & 786 & 1.015 & 79 & 778 & NA \\
\hline
\textbf{C-fat200-1} & 200 & 188 & 188 & 1 & 4 & 188 & 10\\
\hline
\textbf{C-fat200-2} & 200 & 176 & 176 & 1 & 7 & 176 & 10\\
\hline
\textbf{C-fat200-5} & 200 & 142 & 142 & 1 & 6 & 142 & 10\\
\hline
\textbf{C-fat500-1} & 500 & 486 & 486 & 1 & 91 & 486 & NA\\
\hline
\textbf{C-fat500-10} & 500 & 374 & 374 & 1 & 153 & 374 & NA\\
\hline
\textbf{C1000.9} & 1000 & 932 & 948 & 1.017 & 19 & 932 & 5820\\
\hline
\textbf{C2000.9} & 2000 & 1922 & 1946 & 1.012 & 160 & 1922 & 369330\\
\hline
\textbf{Frb30-15-1} & 450 & 420 & 429 & 1.021 & 7 & 420 & 80\\
\hline
\textbf{Frb30-15-2} & 450 & 420 & 431 & 1.026 & 6 & 420 & 100\\
\hline
\textbf{Frb30-15-3} & 450 & 420 & 429 & 1.021 & 5 & 420 & 400\\
\hline
\textbf{Frb30-15-4} & 450 & 420 & 430 & 1.023 & 6 & 420 & 80\\
\hline
\textbf{Frb30-15-5} & 450 & 420 & 428 & 1.019 & 4 & 420 & 170\\
\hline
\textbf{Gen200-p0.9-44} & 200 & 156 & 169 & 1.083 & 1 & 156 & 100\\
\hline
\textbf{Graph50-01} & 50 & 30 & 30 & 1 & 1 & NA & NA \\
\hline
\textbf{Graph50-09} & 50 & 30 & 40 & 1.333 & 1 & NA & NA \\
\hline
\textbf{Graph50-10} & 50 & 30 & 35 & 1.166 & 1 & NA & NA \\
\hline
\textbf{Hamming6-2} & 64 & 32 & 32 & 1 & 1 & 32 & 10\\
\hline
\textbf{Hamming8-2} & 256 & 128 & 128 & 1 & 1 & 128 & 10\\
\hline
\textbf{Hamming10-2} & 1024 & 512 & 512 & 1 & 10 & 512 & 10\\
\hline
\textbf{Johnson8-2-4} & 28 & 24 & 24 & 1 & 3 & 24 & 24 \\
\hline
\textbf{Johnson8-4-4} & 70 & 56 & 56 & 1 & 5 & 56 & NA\\
\hline
\textbf{Johnson16-2-4} & 120 & 112 & 112 & 1 & 5 & 112 & 297\\
\hline
\textbf{Johnson32-2-4} & 496 & 480 & 480 & 1 & 56 & 480 & 920\\
\hline
\textbf{Keller4} & 171 & 160 & 162 & 1.012 & 2 & 160 & NA\\
\hline
\textbf{Keller5} & 776 & 749 & 760 & 1.014 & 106 & 749 & 5127\\
\hline
\textbf{MANN\_a81} & 3321 & 2221 & 2241 & 1.009 & 141 & 2223 & 7100\\
\hline
\textbf{San200-0.7-1} & 200 & 170 & 185 & 1.088 & 3 & 170 & 10\\
\hline
\textbf{San200-0.7-2} & 200 & 182 & 188 & 1.032 & 3 & 182 & 10\\
\hline
\textbf{San400-0.9-1400} & 400 & 300 & 350 & 1.166 & 8 & NA & NA\\
\hline
\textbf{San1000} & 1000 & 985 & 992 &1.007 & 1829 & 985 & 3910\\
\hline
\textbf{Sanr200-0.7} & 200 & 183 & 184 & 1.005 & 1 & 183 & 10\\
\hline
\textbf{Sanr200-0.9} & 200 & 158 & 162 & 1.025 & 1 & 158 & 10\\
\hline
\textbf{Sanr400-0.5} & 400 & 387 & 390 & 1.007 & 8 & 287 & 60\\
\hline
\textbf{Sanr400-0.7} & 400 & 379 & 384 & 1.013 & 9 & 379 & 30\\
\hline
\end{tabular}
\label{result table}
\caption{Obtained results of NEC on a certain set of graph dataset \(DIMACS \)}
\end{table}

\pagebreak

\section{conclusions}

NEC proves to be extremely efficient by calculating optimal or near optimal vertex cover on known benchmark graphs. Its worst case has a selection ratio of 1.5[7].
It is experimentally observed to be fast (as compared to current algorithms such as COVER) while the candidate node selection makes sure that it selects high valued candidates. One area left to explore is the impact of graph density and selection ratio and should be taken up in future studies.

\end{document}